\documentclass{article}
\PassOptionsToPackage{numbers,compress}{natbib}
\usepackage[final]{template22}
\usepackage{float} 
\usepackage[utf8]{inputenc} % allow utf-8 input
\usepackage[T1]{fontenc}    % use 8-bit T1 fonts
\usepackage{microtype}
\usepackage{times}
\usepackage{graphicx}
\usepackage{amsmath,amssymb,mathbbol}
\usepackage{acronym}
\usepackage{enumitem}
\usepackage[pagebackref=true,breaklinks=true,colorlinks]{hyperref}
\usepackage{balance}
\usepackage{xspace}
\usepackage{setspace}
\usepackage[skip=3pt,font=small]{subcaption}
\usepackage[skip=3pt,font=small]{caption}
\usepackage[dvipsnames]{xcolor}
\usepackage[capitalise]{cleveref}
\usepackage{booktabs,tabularx,colortbl,multirow,array,makecell}

\usepackage{fancyhdr}
\hypersetup{pdfencoding=auto,colorlinks=true,allcolors=black}

\makeatletter
\DeclareRobustCommand\onedot{\futurelet\@let@token\@onedot}
\def\@onedot{\ifx\@let@token.\else.\null\fi\xspace}
\def\eg{\emph{e.g}\onedot} 

\def\ie{\emph{i.e}\onedot}

\makeatother

\definecolor{gray}{gray}{0.9}

\makeatletter
\renewcommand{\paragraph}{%
  \@startsection{paragraph}{4}%
  {\z@}{0ex \@plus 0ex \@minus 0ex}{-1em}%
  {\hskip\parindent\normalfont\normalsize\bfseries}%
}
\makeatother

\graphicspath{{figures/}}

\crefname{algorithm}{Alg.}{Algs.}
\Crefname{algorithm}{Algorithm}{Algorithms}
\crefname{section}{Sec.}{Secs.}
\Crefname{section}{Section}{Sections}
\crefname{table}{Tab.}{Tabs.}
\Crefname{table}{Table}{Tables}
\crefname{figure}{Fig.}{Fig.}
\Crefname{figure}{Figure}{Figure}

\acrodef{pku}[PKU]{Peking University}

\title{Seismic Phase Picking}

\author{%
  Ruihuan Wang \\
  School of Electronics Engineering and Computer Science \\
  Peking University\\
  \texttt{dylanwr@stu.pku.edu.cn} \\
  \And
  Yuchen Wang \\
  School of Electronics Engineering and Computer Science \\
  Peking University\\
  \texttt{2100013153@stu.pku.edu.cn} \\
}

\begin{document}
\maketitle

\begin{abstract}
Seismic phase picking, which aims to determine the arrival time of P- and S-waves according to seismic waveforms, is fundamental to earthquake monitoring. Generally, manual phase picking is trustworthy, but with the increasing number of worldwide stations and seismic monitors, it becomes more challenging for human to complete the task comprehensively. In this work, we explore multiple ways to do automatic phase picking, including traditional and learning-based methods. 
\end{abstract}

\section{Introduction}

Seismic phase picking and detection are the first major steps for earthquake localization and further seismology research. With an increasing number of seismic monitors and waveform data recorded by them, more efficient and robust tools are in urgent need.

For automatic phase picking, there are two mainstream types of method. One is traditional method\citep{akazawa2004technique, allen1978automatic, takanami1991estimation}, which apply well designed filters or matrix to the waveform arrays. Another is learning-based methods\citep{mousavi2020earthquake, ronneberger2015u, zhu2019phasenet}, which build machine learning models for the picking task. Network for multi-station phase picking\citep{chen2022cubenet, feng2022edgephase} also appear in recent years.

In this work, we try both traditional and learning-based models to do phase picking and analyze their performance. 
Due to time and computing resource constraint, out work mainly focuses on single-trace phase picking and some of our ideas are not tested, but we also propose them in \cref{sec:discussion}.

\section{Dataset}

We use the INSTANCE dataset\citep{michelini2021instance} for training and testing.

\subsection{Basic information}

The INSTANCE dataset consists of nearly 1.2 million 3-component waveform traces from about 50,000 earthquakes and more than 130,000 noise 3-component waveform traces. The waveform dataset is accompanied by metadata consisting of more than 100 parameters providing comprehensive information, and thus is suitable for machine learning analysis.

\subsection{Features}

\begin{figure}[ht!]
     \centering
     \begin{subfigure}[b]{0.48\linewidth}
        \centering
        \includegraphics[width=\linewidth]{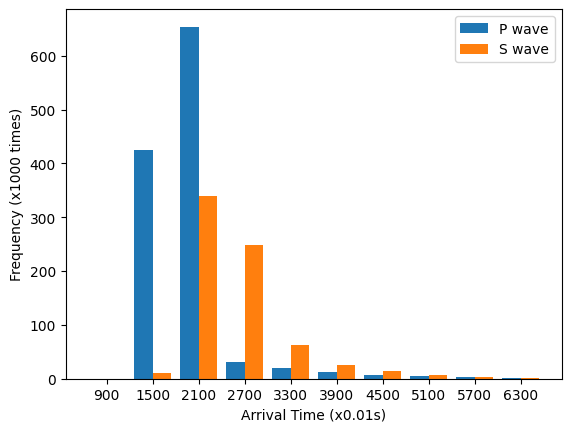}
        \caption{Distribution of arrival time of P- and S-waves. Some of the waveform traces do not contain S-waves. \quad \quad}
        \label{fig:sub1}
     \end{subfigure}%
     \hfill
     %\begin{subfigure}[b]{0.48\linewidth}
     %   \centering
     %   \includegraphics[width=\linewidth]{images/deltatStat.png}
     %   \caption{Distribution of $\Delta t = t_\mathrm{S-wave} - t_\mathrm{P-wave}$. Consider only waveforms that contain both P- and S-waves.}
     %   \label{fig:sub2}
     %\end{subfigure}
     %\quad
    \begin{subfigure}[b]{0.48\linewidth}
        \centering
        \includegraphics[width=\linewidth]{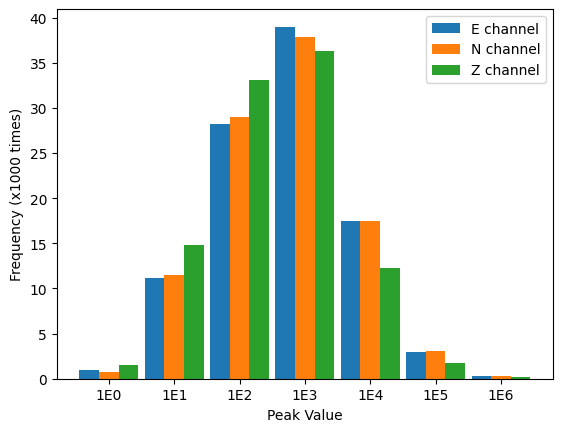}
        \caption{Distribution of peak values in three channels (of 100,000 samples). Note that the value on x-axis is increasing exponentially.}
        \label{fig:sub2}
    \end{subfigure}
    \caption{A glance into distributional features of the INSTANCE dataset.}
     \label{fig:features}
\end{figure}

In this work, we mainly focus on parameters \texttt{trace\_P\_arrival\_sample} and \texttt{trace\_S\_arrival\_sample}, which represent the arrival time of P- and S-waves of an earthquake. 
\cref{fig:features} shows some distributional features of the dataset. In \cref{sec:datapreprocessing}, we will further discuss how the features influence our data building work.

\section{Traditional method}

A classical traditional phase picking algorithm is AR-AIC model, which is widely used in related works (\eg \citep{akazawa2004technique, takanami1988new}). We apply AR-pick tool from ObsPy\citep{beyreuther2010obspy} to 1,000 samples from INSTANCE and \cref{tab:traditionalresult} shows our result.

\begin{table}[ht!]
    \caption{\textbf{Result of AR pick.} $\mu$ and $\sigma$ are mean and standard deviation of errors (\ie ground truth - prediction) in seconds respectively. Pr is precision. Arrival time residuals that are less than threshold are counted as true positive.}
    \label{tab:traditionalresult}
    \centering
    \begin{tabular}{c|cc|cc}
        \toprule
        Phase   &  $\mu$    & $\sigma$      & threshold (s)     & Pr        \\
        \midrule
                &           &               &     0.10          &   0.34    \\
                &           &               &     0.30          &   0.63    \\
          P     &  0.86     &    8.84       &     0.50          &   0.71    \\
                &           &               &     1.00          &   0.77    \\
                &           &               &     2.00          &   0.80    \\
        \midrule
                &           &               &     0.10          &   0.13    \\
                &           &               &     0.30          &   0.28    \\
          S     &   -4.17   &    17.20      &     0.50          &   0.40    \\
                &           &               &     1.00          &   0.64    \\
                &           &               &     2.00          &   0.84    \\
        \bottomrule
    \end{tabular}
\end{table}

Generally, the performance of AR pick is not satisfying enough. (1) \cref{fig:sub1} shows that for most records, the arrival time are less than 40 seconds, but the mean of error can reach over 4 seconds for S-wave, which is an error of around 10\%. (2) The standard deviation is relatively large, which demonstrates a lack of robustness. (3) Even when we set a threshold to 2s, Pr value still cannot reach 0.90. 

Therefore, AR Pick is not capable of precise phase picking and cannot replace manual phase picking. A more powerful method is needed. 

\section{Learning-based method}

\subsection{Data preprocessing and label designing}
\label{sec:datapreprocessing}

The first task to do supervised learning is to build data-label pairs.

\begin{figure}[ht!]
     \centering
     \begin{subfigure}[b]{0.48\linewidth}
         \centering
        \includegraphics[width=\linewidth]{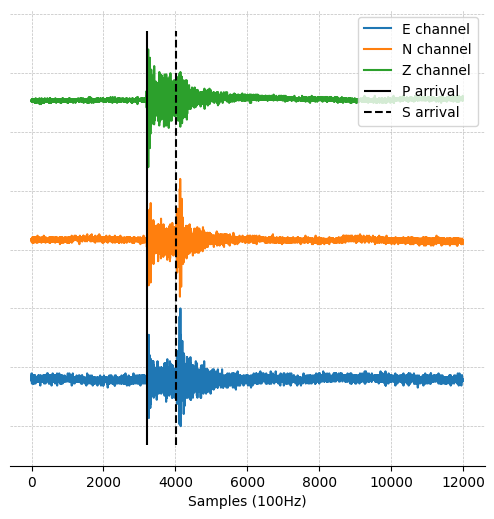}
        \caption{A sample from INSTANCE.}
        \label{fig:sample}
     \end{subfigure}
        \hfill
    \begin{subfigure}[b]{0.48\linewidth}
        \centering
        \includegraphics[width=\linewidth]{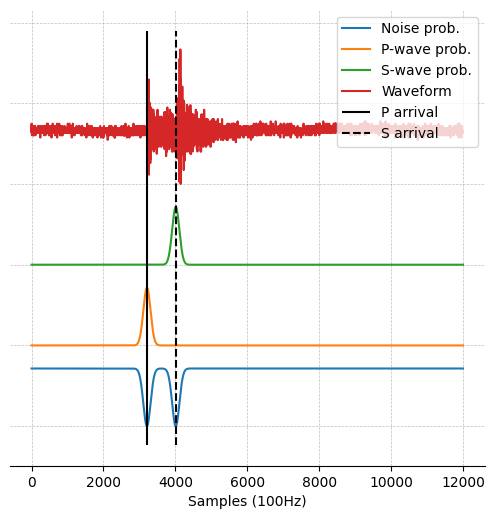}
        \caption{A labeled example (using Gaussian filter).}
        \label{fig:label}
    \end{subfigure}
    \caption{An example of data and labeling.}
\end{figure}

As shown in \cref{fig:sample}, the data from INSTANCE have a uniform length of 12,000 samples (120 seconds with 100Hz sampling rate). But according to \cref{fig:sub1}, arrival time of signals are mainly distributed in the range of 1,000 - 6,000. So it is reasonable to set a waveform window for training, which means to take the former 6,000 record samples and drop the rest. In this way we are able to accelerate the training progress while taking full advantage of the dataset.

Besides, from \cref{fig:sub2}, we see a relatively wide range of peak value distribution. So it is not wise to directly put raw data into the net. We normalize \textit{each component} by 
\begin{equation*}
    \begin{aligned}
    \mathrm{Unbiased}_i &= x_i - \overline{x_i}, \\
    \mathrm{Normed}_i &= \cfrac{ \mathrm{Unbiased}_i}{\max  \left\{\left|\mathrm{Unbiased}_i\right|\right\}}
    \end{aligned}
\end{equation*}
where $i \in$ \{E, N, Z\} represents a certain channel and $x_i$ is the raw value of the $i$-th component.

As for labels, an intuitive idea is to treat the phase picking task as a regression problem, and label each waveform with 2 numbers (arrival time of P- and S- wave). But there are some challenges if using this method. (1) Not all traces contain both types of wave. Sometimes the station fail to record S-wave because S-wave cannot travel through the molten outer core of the Earth. Besides, when coping with pure noise waveform, such labeling seems impossible to be done. (2) The arrival time has uncertainty and it is improper to see the arrivals as single numbers, so we expect the ground truth will be centered on the manual picks with some uncertainty.

For this reason, we apply a mask which reach peak value at the manual picked time point and the nearby data points have reduced values (\cref{fig:label} shows an example labeled by Gaussian filters. We also try triangle and box filters in our experiment), representing the probability of arrival. Data points other than arrival of P- and S-waves are considered noise. Thus the probability distribution of noise can be calculated by
$$\mathrm{Prob_{noise} = 1-Prob_{P}-Prob_{S}},$$
where $\mathrm{Prob}$ means the probability of each class.

\subsection{Models}

\subsubsection{CNN-based model: PhaseNet}

\begin{figure}[ht!]
    \centering
    \includegraphics[width=0.75\linewidth]{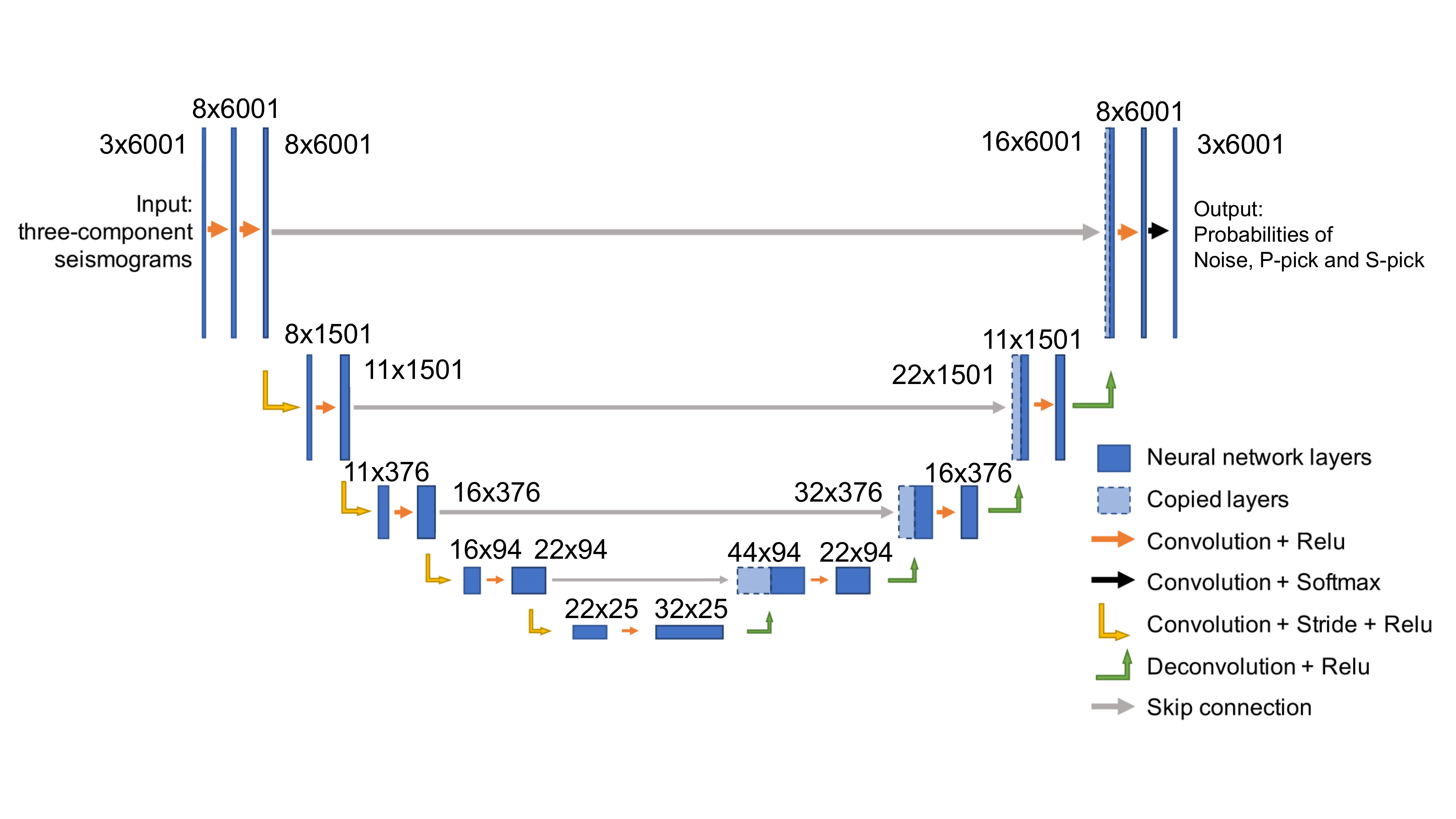}
    \caption{PhaseNet architecture (layer dimensions modified by us) from \citep{zhu2019phasenet}.}
    \label{fig:phasenet}
\end{figure}

The architecture of PhaseNet (\cref{fig:phasenet}) is modified from U-net\citep{ronneberger2015u}, with 4 down-sampling and 4 up-sampling stages. 1-D convolutions and ReLU activations are applied inside each stage. A skip connection that directly concatenates the left output to the right layer is also used to improve convergence. In the last layer, the softmax function is used to calculate probabilities of noise, P-wave and S-wave.

\subsubsection{Attention-based model: EQTransformer}

\begin{figure}[ht!]
    \centering
    \includegraphics[width=0.9\linewidth]{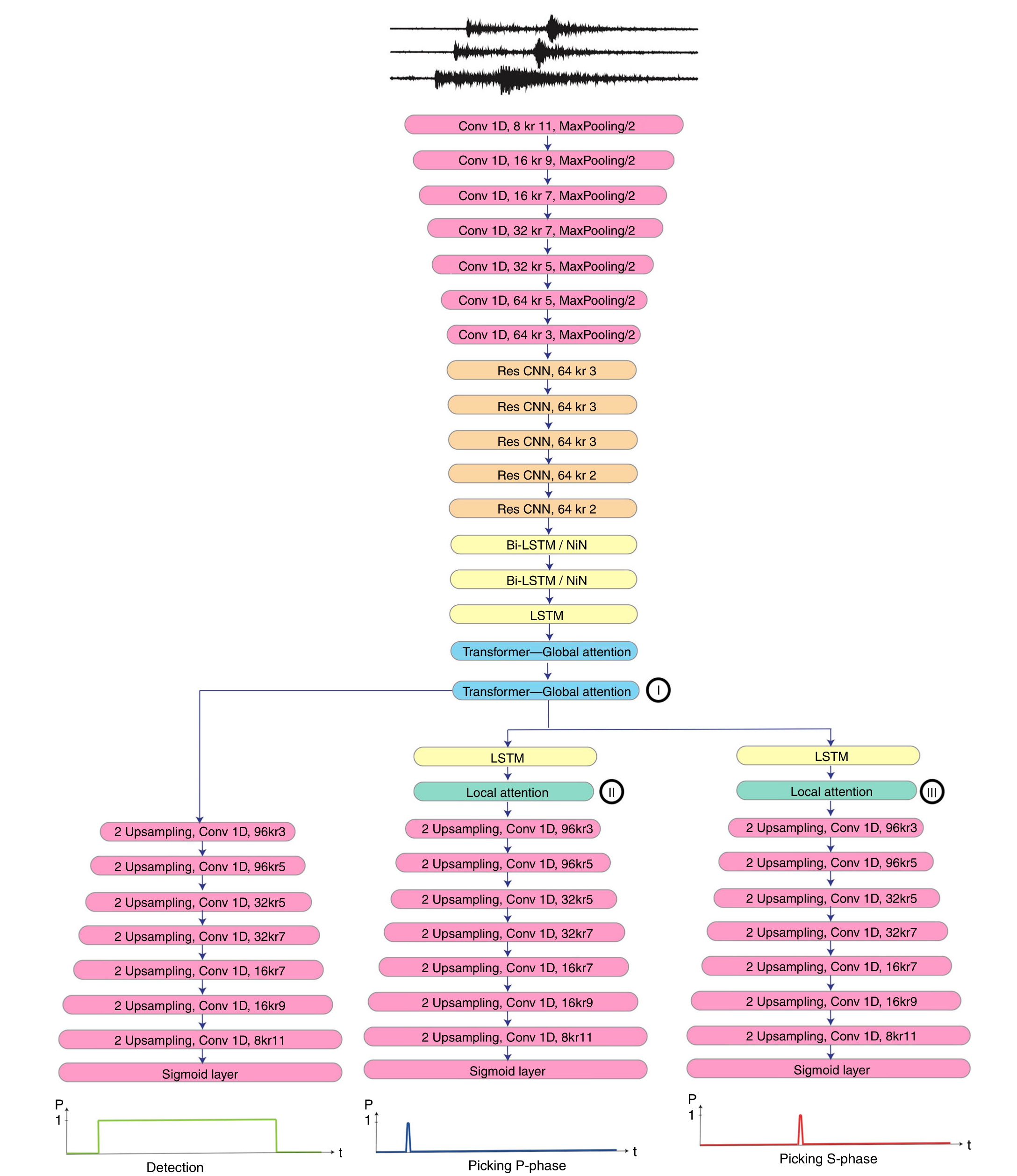}
    \caption{EQTransformer architecture (taken from \citep{mousavi2020earthquake}).}
    \label{fig:eqt}
\end{figure}

As shown in \cref{fig:eqt}, EQTransformer consists of a deep encoder and three separate decoders that detect earthquake, pick P- and S-wave respectively. 

\subsection{Loss}

The loss function is defined as cross-entropy between the ground truth distribution ($y$) and predicted distribution ($y'$) to measure the divergence between them:
$$
\mathcal{L}(y, y') = -\sum_{i=1}^{3}\sum_{\mbox{sample points}}y_{i}(x)\log (y'_i(x) + \varepsilon),
$$
where $i=1,2,3$ represents noise, P- and S-wave. $\varepsilon$ is added to avoid taking the logarithm of 0 and is set to 1E-5 in our experiment.

\section{Experiment}

The experiment progress does not go as smooth as expected.

We first try to implement and train PhaseNet. Since it is originally designed to cope with waveform with a length of 3,000, we do a dimension modification first (see \cref{fig:phasenet}). However, it is strange that the model does not converge correctly. We manage to print out the predicted result and find that all values of probabilities of P- and S-waves are 0, which means the model seems to regard all sample points as noise. We have made lots of efforts to solve this, such as adding/removing layers (like Dropout and BatchNormalization), trying different strategies to preprocess the data, trying different sets of kernel sizes, increase the weights of wave layers when computing loss, etc. Yet eventually the problem remains unsolved.

There is also a possibility that there are bugs in our training progress due to our lack of deep learning practical experience on such a large dataset. 

As for EQTransformer, we also implement a PyTorch version and manage to fit the pretrained model into our torch version implementation. Though not trained on INSTANCE, the result (see \cref{tab:cmp}) still beat AR Pick on this dataset, which shows a high generalization and reliability of learning-based picker.

\begin{table}[ht!]
    \caption{\textbf{Comparison of EQTransformer and AR Pick.} $\mu$ and $\sigma$ are mean and standard deviation of errors (\ie ground truth - prediction) in seconds respectively. Pr, Re, F1 are precision, recall and F1-score. Arrival time residuals that are less than $\Delta t$ = 0.5s are counted as true positive. When using AR Pick, we simply enforce it to predict the arrival time of both P- and S-wave, no matter whether there is corresponding wave. So Re and F1 of AR Pick are meaningless here.}
    \label{tab:cmp}
    \centering
    \begin{tabular}{c|c|cc|ccc}
    \toprule
    Phase & Model &  $\mu$    & $\sigma$      &   Pr      & Re   & F1 \\
    \midrule
    \multirow{2}{*}{P}   & AR Pick & 0.86 & 8.84 & 0.71 & - & - \\
    & EQTransformer  & -0.27 & 4.18 & 0.76 & 1.00 & 0.86 \\
    \midrule
    \multirow{2}{*}{S}   & AR Pick & -4.17 & 17.20 & 0.40 & - & - \\
    & EQTransformer  & -0.39 & 5.37 & 0.61 & 0.88 & 0.72 \\
    \bottomrule
    \end{tabular}
\end{table}

\newpage
\cref{fig:res} shows some visualized results of EQTransformer and AR Pick. In each figure, only the E channel of waveform is shown. We can clearly understand how noise influences the prediction. When there is little noise (\cref{fig:res1} and \cref{fig:res2}), both models work well. When noise gets stronger (\cref{fig:res3} and \cref{fig:res4}), AR Pick may fail to give a correct prediction. If the amplitude of noise is very close to the waves (\cref{fig:res5} and \cref{fig:res6}), both models cannot deal with the condition very well, and it is also challenging for humans to pick the phases out if only given single trace. This is the reason why we need multi-trace based phase picking techniques. Through analyzing the waveforms of the same earthquake recorded by multiple stations, the models are able to get more robust to noise.

\begin{figure}[H]
     \begin{subfigure}[b]{0.98\linewidth}
        \centering
        \includegraphics[width=\linewidth]{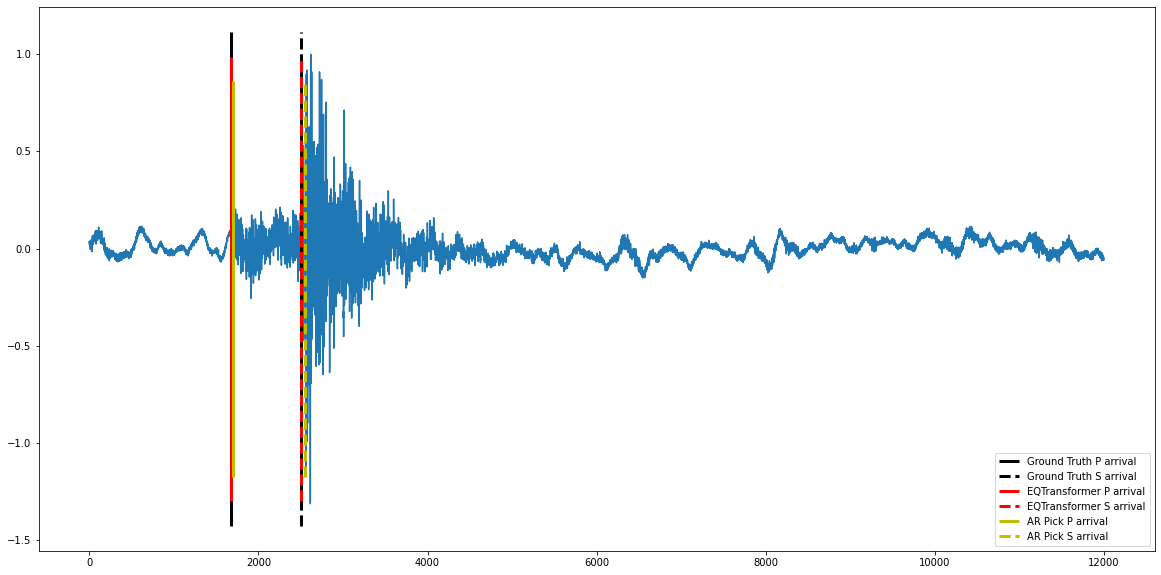}
        \caption{Slight noise, both models work well.}
        \label{fig:res1}
     \end{subfigure}%
     \quad
    \begin{subfigure}[b]{0.98\linewidth}
        \centering
        \includegraphics[width=\linewidth]{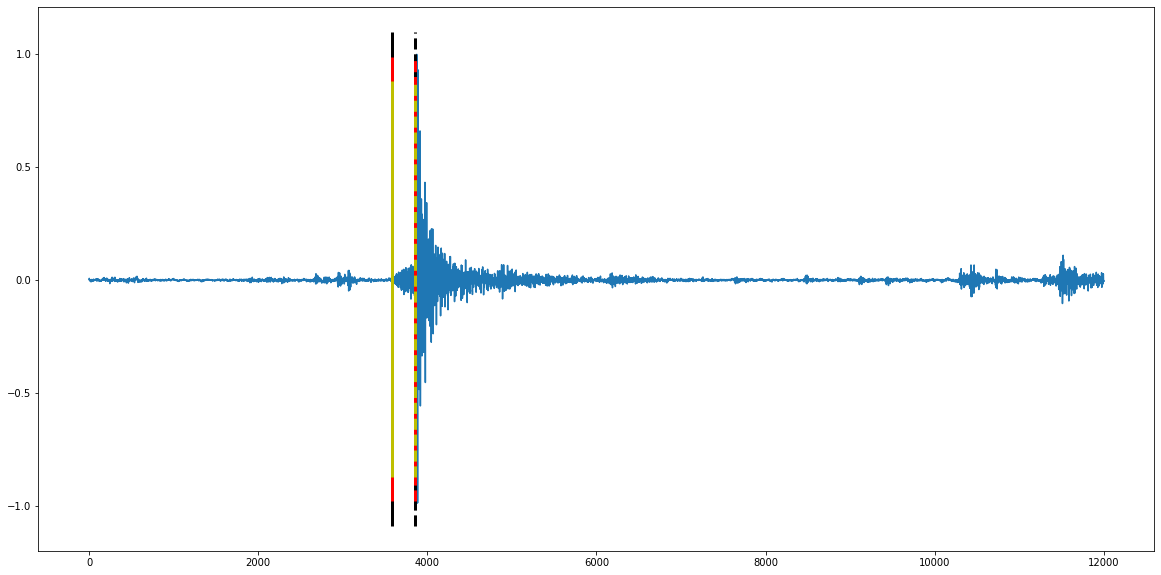}
        \caption{Almost no noise, both models work well}
        \label{fig:res2}
    \end{subfigure}
    \quad
    \begin{subfigure}[b]{0.98\linewidth}
        \centering
        \includegraphics[width=\linewidth]{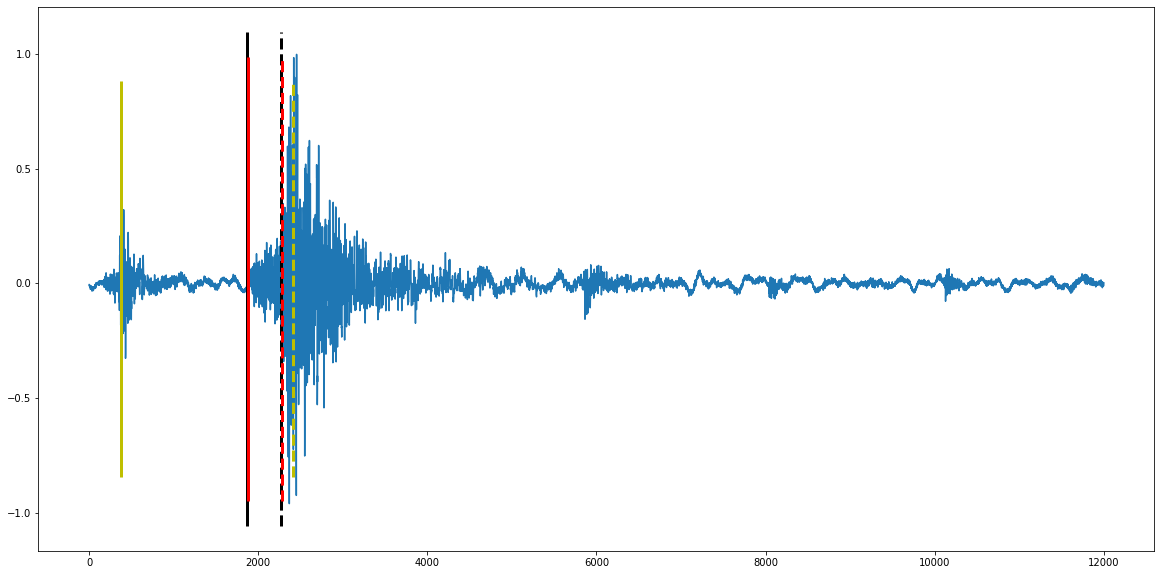}
        \caption{AR Pick cheated by noise.}
        \label{fig:res3}
    \end{subfigure}
\end{figure}
\begin{figure}[H]
    \ContinuedFloat
    \setcounter{subfigure}{3}
    \begin{subfigure}[b]{0.98\linewidth}
        \centering
        \includegraphics[width=\linewidth]{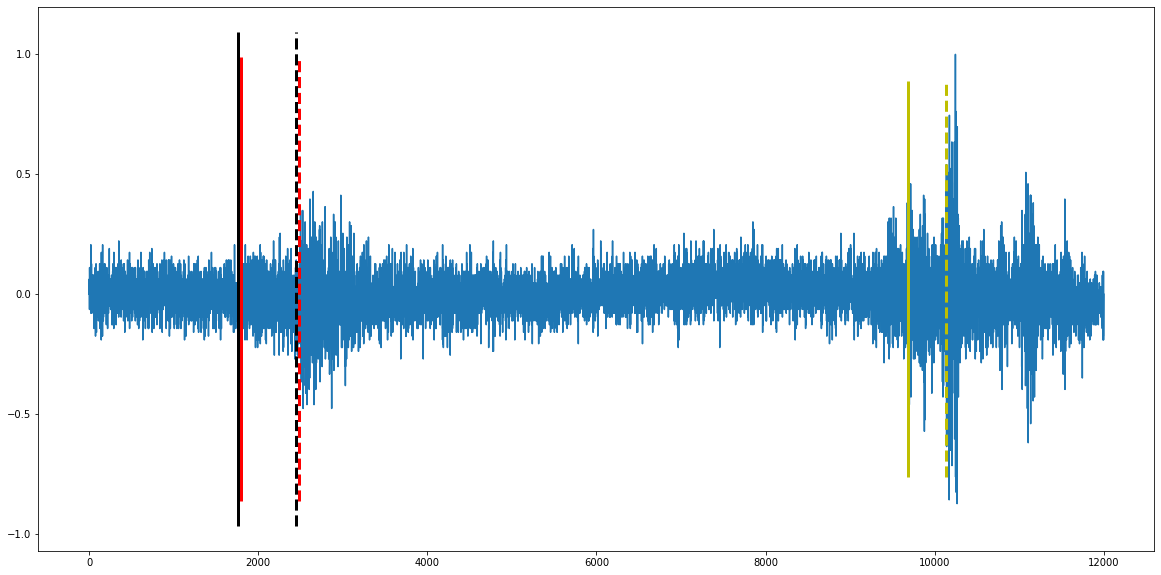}
        \caption{AR Pick cheated by noise.}
        \label{fig:res4}
    \end{subfigure}
    \quad
    \begin{subfigure}[b]{0.98\linewidth}
        \centering
        \includegraphics[width=\linewidth]{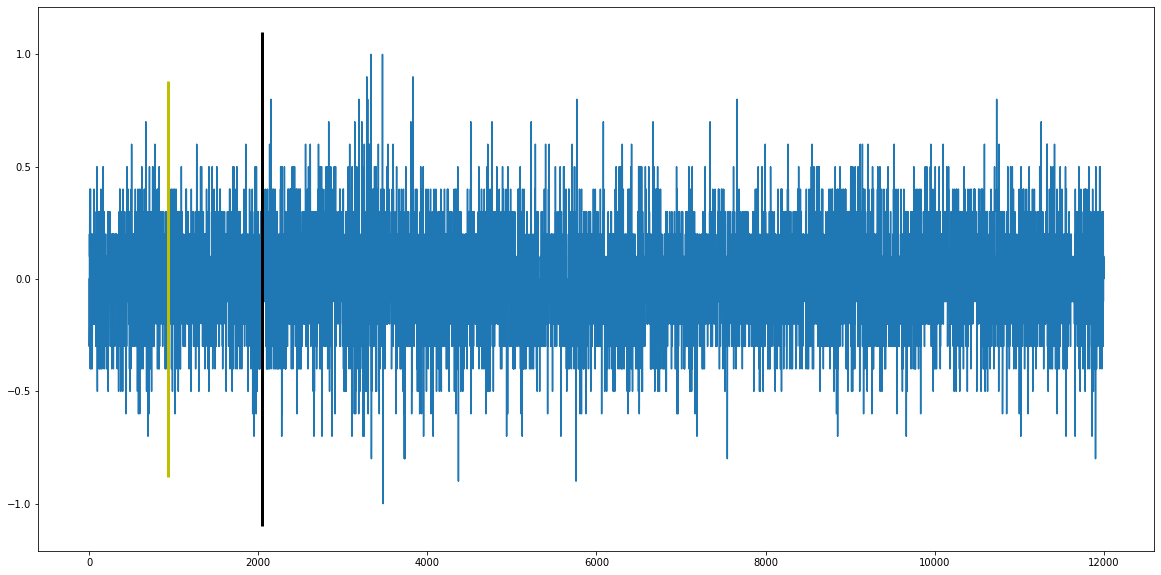}
        \caption{Strong noise, EQTransformer cannot find waves and AR Pick gives a wrong prediction.}
        \label{fig:res5}
    \end{subfigure}
    \quad
    \begin{subfigure}[b]{0.98\linewidth}
        \centering
        \includegraphics[width=\linewidth]{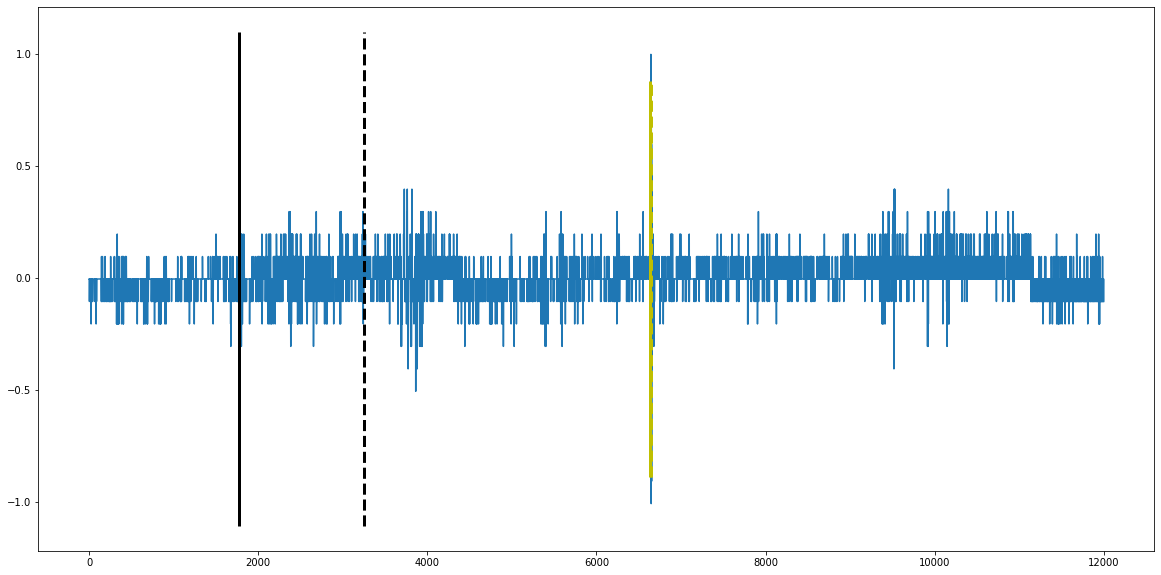}
        \caption{Distribution of peak values in three channels (of 100,000 samples). Note that the value on x-axis is increasing exponentially.}
        \label{fig:res6}
    \end{subfigure}
    \caption{Strong noise, EQTransformer cannot find waves and AR Pick gives wrong predictions.}
    \label{fig:res}
\end{figure}

\section{Discussion}
\label{sec:discussion}

Some of our ideas to solve the phase picking problem will be introduced in this section. Since the attempts to fix and finetune PhaseNet takes up too much time and effort, these ideas still remain untried and untested.

\subsection{Ideas on single-trace phase picking}

Transformer and Attention\citep{NIPS2017_3f5ee243} has been widely used in sequence processing and even image classification\citep{dosovitskiy2020image}. EQTransformer and some other recent models\citep{liao2021arru,liao2022red} also introduces several Attention blocks into their net. Transformer is proved to be a strong tool, and we notice that in many researching fields, convolution-free and Transformer-only architectures have been raised and can rank top\citep{cao2021swin, 2205.13425, he2022transfg, wang2021pyramid}. We do not find similar works in the field of seismic phase picking and we wonder what a pure transformer net will bring to this task.

\subsection{Ideas on multi-trace phase picking}

An important concern is that in manual phase picking process, humans would use waveform data from multiple stations to determine
whether an ambiguous case should be mapped to a seismic phase. As we learned from the EdgePhase\cite{feng2022edgephase}, we may use Graph to illustrate the relation between stations, like, determine stations as neighbors if their distance is less than 100 km, and if neighbors consistently get a noise-like wave, the probabilities that it may be from a seismic phase are higher. Also we need to use single trace network to get the \textit{ambiguous} information about the wave we detect.Then we can integrate the information in the nearby area to a more precise and reliable summary.

\bibliographystyle{plainnat}
\bibliography{reference}

\end{document}